\documentclass[11pt]{article}
\usepackage{graphicx}
\usepackage[T1]{fontenc}
\usepackage{textcomp}
\usepackage{natbib,float}
\usepackage{titlesec,color}
\usepackage{ctable,colortbl,color,array}
\usepackage[table]{xcolor}
\usepackage{setspace,hyperref,titlesec,mathpazo}
\usepackage[top=1in, bottom=1in, left=1in, right=1in]{geometry}

\bibpunct{(}{)}{;}{a}{}{,}

\begin{document}

\newcommand{\Keywords}[1]{\par\noindent 
{\small{\bf Keywords\/}: #1}}

\newcommand{\RunningTitle}[1]{\par\noindent 
{\small{\bf Running Title\/}: #1}}

\title{\emph{De novo} genomic analyses for non-model organisms: an evaluation of methods across a multi-species data set}
\author{Sonal Singhal \\ \\
singhal@berkeley.edu \\ \\
Museum of Vertebrate Zoology \\ University of California, Berkeley \\ 3101 Valley Life Sciences Building \\ Berkeley, California 94720-3160 \\ \\
Department of Integrative Biology \\ University of California, Berkeley \\ 1005 Valley Life Sciences Building \\ Berkeley, California 94720-3140 \\ \\
}
\maketitle

\begin{abstract}
High-throughput sequencing (HTS) is revolutionizing biological research by enabling scientists to quickly and cheaply query variation at a genomic scale. Despite the increasing ease of obtaining such data, using these data effectively still poses notable challenges, especially for those working with organisms without a high-quality reference genome.  For every stage of analysis -- from assembly to annotation to variant discovery -- researchers have to distinguish technical artifacts from the biological realities of their data before they can make inference. In this work, I explore these challenges by generating a large \emph{de novo} comparative transcriptomic dataset data for a clade of lizards and constructing a pipeline to analyze these data. Then, using a combination of novel metrics and an externally validated variant data set, I test the efficacy of my approach, identify areas of improvement, and propose ways to minimize these errors. I find that with careful data curation, HTS can be a powerful tool for generating genomic data for non-model organisms. \\ 
\RunningTitle{\emph{De novo} genomic analyses for non-model organisms}
\end{abstract}

\onehalfspace

\section{Introduction}
High-throughput sequencing (HTS) is poised to revolutionize the field of evolutionary genetics by enabling researchers to assay thousands of loci for organisms across the tree of life. Already, HTS data sets have facilitated a wide range of studies, including identification of genes under natural selection \citep{Yi2010}, reconstructions of demographic history \citep{Luca2011}, and broad scale inference of phylogeny \citep{Smith2011}. Daily, sequencing technologies and the corresponding bioinformatics tools improve, making these approaches even more accessible to a wide range of researchers. Still, acquiring HTS data for non-model organisms is non-trivial, especially as most applications were designed and tested using data for organisms with high-quality reference genomes. Assembly, annotation, variant discovery, and homolog identification are challenging propositions in any genomics study \citep{Baker2012,Nielsen2011}; doing the same \emph{de novo} for non-model organisms adds an additional layer of complexity. Already, many studies have collected HTS data sets for organisms of evolutionary and ecological interest \citep{Hohenlohe2010,Keller2012,Ellegren2012} and have developed associated pipelines. Some have published these pipelines to share with other researchers \citep{Catchen2011,Hird2011,DeWit2012}; such programs make HTS more accessible to a wider audience and serve as an excellent launching pad for beginning data analysis. However, because each HTS data set likely poses its own challenges and idiosyncracies, researchers must evaluate the efficacy and accuracy of any pipeline for their data sets before they are used for biological inference. Evaluating pipeline success is easier for model organisms, where reference genomes and single nucleotide polymorphism (SNP) sets are more common; however, for most non-model organisms, we often lack easy metrics for gauging pipeline efficacy. 

In this study, I generate a large HTS data set for five individuals each from seven phylogeographic lineages in three species of Australian skinks (family: \emph{Scincidae}; Fig. S2), for which the closest assembled genome (\emph{Anolis carolinesis}) is highly divergent (most recent common ancestor [MRCA], 150 million years ago [Mya], \cite{Alfoldi2011}). These seven lineages are closely related; they shared a MRCA about 25 Mya \citep{Skinner2011}. This clade is the focus of a set of studies looking at introgression across lineage boundaries \citep{Singhal2012}, and to set the foundation for this work, I generate and analyze genomic data for lineages meeting in four of these contacts, two of which are between sister-lineages exhibiting deep divergence (\emph{Carlia rubrigularis} N/S, \emph{Lampropholis coggeri} C/S) and two which show shallow divergence (\emph{Saproscincus basiliscus} C/S, \emph{Lampropholis coggeri} N/C) (Fig. S2). I use these data to develop a bioinformatics pipeline to assemble and annotate contigs, and then, to define variants within and between lineages and identify homologs between lineages. Using both novel and existing metrics and an externally validated SNP data set, I am able to test the effectiveness of this pipeline across all seven lineages. In doing so, I refine my pipeline, identify remaining challenges, and evaluate the consequences of these challenges for downstream inferences. My work makes suggestions to other researchers conducting genomics research with non-model organisms, offers ideas on how to evaluate the efficacy of pipelines, and discusses how the technical aspects of HTS sequencing can affect biological inference. 

\section{Methods}
All bioinformatic pipelines are available as {\sffamily Perl} scripts on \url{https://sites.google.com/site/mvzseq/original-scripts-and-pipelines/pipelines}, and they are summarized graphically in Figs. 1A and S1. I have also shared {\sffamily R} scripts \citep{CRAN} that use {\sffamily ggplot2} to do the statistical analyses and graphing presented in this paper \citep{ggplot2}.
\subsection{Library Preparation and Sequencing}
Even though costs of sequencing continue to drop and assembly methods improve \citep{Glenn2011,Schatz2010}, whole-genome \emph{de novo} sequencing remains inaccessible for researchers interested in organisms with large genomes (\emph{i.e.}, over 500 Mb) and for researchers who wish to sample variation at the population level. Thus, most \emph{de novo} sequencing projects must still use some form of complexity reduction (\emph{i.e.}, target-based capture or restriction-based approaches) in order to interrogate a manageable portion of the genome. Here, I chose to sequence the transcriptome, because it is appropriately sized to ensure high coverage and successful \emph{de novo} assembly, I will surely obtain homologous contigs across taxa, I can capture both functional and non-coding variation, and assembly can be validated by comparing to known protein-coding genes.

Liver and, where appropriate, testes samples were collected from adult male and female lizards during a field trip to Australia in fall 2010 (Table S1); tissues and specimens are accessioned at the Museum of Vertebrate Zoology, UC-Berkeley. I extracted total RNA from RNA-later preserved liver tissues using the Promega Total RNA SV Isolation kit. After checking RNA quality and quantity with a Bioanalyzer, I used the Illumina mRNA TruSeq kit to prepare individually barcoded cDNA libraries. Final libraries were quantified using qPCR, pooled at equimolar concentrations, and sequenced using four lanes of 100bp paired-end technology on the Illumina HiSeq2000.
\subsection{Data Quality and Filtration}
I evaluated raw data quality by using the {\sffamily FastQC} module \citep{Andrews} and in-house Perl scripts that calculate sequencing error rate. Sequencing error rates for Illumina reads have been reported to be as high as 1\% \citep{Minoche2011}; such high rates can both lead to poor assembly quality and false positive calls for SNPs. To compare to these reported values, I derived an empirical estimate of sequencing error rate. To do so, I aligned a random subsample of overlapping forward-reverse reads (N=100,000) using the local aligner {\sffamily blat} \citep{Kent2002}, identified mismatches and gaps, and calculated error rates as the total number of errors divided by double the length of aligned regions. Data were then cleaned: exact duplicates due to PCR amplification were removed, low-complexity reads (e.g., reads that consisted of homopolymer tracts or more than 20\% '{\sffamily N}'s) were removed, reads were trimmed for adaptor sequence and for quality using a sliding window approach implemented in {\sffamily Trimmomatic} \citep{Lohse2012}, reads matching contaminant sources (e.g., ribosomal RNA and human and bacterial sources) were removed via alignment to reference genomes with {\sffamily Bowtie2} \citep{Langmead2012}, and overlapping paired reads were merged using {\sffamily Flash} \citep{Magoc2011}. Following data filtration but prior to read merging, I again estimated sequencing error rates using the method described above.

\subsection{\emph{de novo} Assembly}
Determining what kmer, or nucmer length, to use is key in \emph{de novo} assembly of genomic data \citep{Earl2011}. In assembling data with even coverage, researchers typically use just one kmer \citep{Earl2011}; however, with transcriptome data, contigs have uneven coverage because of gene expression differences \citep{Martin2011}. Thus, some have shown the ideal strategy for transcriptomes is to assemble data at multiple kmers and then assemble across the assemblies to reduce redundancy \citep{SurgetGroba2010}.  To assemble across assemblies, I first identify similar contigs using clustering algorithms ({\sffamily cd-hit-est}; \citep{Li2006}) and local alignments ({\sffamily blat}; \citep{Kent2002}) and then assemble similar contigs using a light-weight \emph{de novo} assembler ({\sffamily cap3}; \citep{Huang1999}). I used this multi-kmer, custom merge approach along with other existing approaches, including:
\begin{itemize}
\item A single kmer approach implemented in the program {\sffamily Trinity} (a \emph{de novo} RNA transcript assembler; \citep{Grabherr2011})
\item A single kmer approach implemented in {\sffamily ABySS} (a \emph{de novo} genomic assembler; \citep{Simpson2009}), {\sffamily Velvet} (a \emph{de novo} genomic assembler; \citep{Zerbino2008}), and {\sffamily SOAPdenovo-Trans} (a \emph{de novo} RNA transcript assembler; \citep{Li2010}), which I implemented as a multi-kmer approach using my custom merge script
\item A multi-kmer approach implemented in the program {\sffamily OASES} \citep{Schulz2012}
\end{itemize}
I explore a wide-range of assembly methods because generating a high-quality and complete assembly is key for almost all downstream applications. Particularly with genome assembly, which is both an art and a science, researchers should try multiple approaches and evaluate their efficacy before further analyses \citep{Earl2011}. However, without a reference genome, evaluating the quality of a \emph{de novo} assembly is challenging. Here, I implement novel metrics for evaluating \emph{de novo} transcriptome assemblies.  In addition to existing metrics in the literature (N50, mean contig length, total assembly length) \citep{Martin2011}, I determined which proportion of reads were used in the assembly, measured putative levels of chimerism in transcripts due to misassemblies, determined the proportion of assembled transcripts that could be annotated and the accuracy of these transcripts (as determined by the number of nonsense mutations), and calculated the completeness and contiguity of the assembly \citep{Martin2011}.

Here, I assembled across all individuals in a lineage rather than assembling each individual separately. Although this introduced additional polymorphism into the data which can reduce assembly efficiency \citep{Vinson2005}, previous work suggests the additional data lead to more complete assemblies (Singhal, unpublished). 

\subsection{Annotation}
Following evaluation of my final assemblies, I chose the best assembly for annotation to protein databases. Determining the most appropriate database for annotation is important, so I tested multiple options, including using a single-species database, whether from a distantly-related but well-annotated genome or closely-related but poorly-annotated genome, using a multi-species database, or using a curated protein set, such as {\sffamily UniRef90} \citep{Susek2007}. For one randomly selected lineage, I tested the efficiency and accuracy of five different reference databases: 
\begin{itemize}
\item the non-redundant {\sffamily Ensembl} protein database \citep{Flicek2012} for the lizard \emph{Anolis carolinensis}; with a most-recent common ancestor to my lineages of about $\approx$150 mya, it is the closest available genome \citep{Alfoldi2011},
\item the non-redundant {\sffamily Ensembl} protein data set for \emph{Gallus gallus}, whose genome is higher quality than the \emph{Anolis} genome but is more distantly related ($\approx$250 mya),
\item a non-redundant, curated data set ({\sffamily UniRef90}) of proteins from a wide range of organisms, whose genes have been clustered at 90\% similarity,
\item a highly-redundant {\sffamily Ensembl} protein data set for eight vertebrates sequenced to high quality (human, dog, rat, mouse, platypus, opossum, dog, chicken),
\item a highly-redundant {\sffamily Ensembl} protein data set for the 54 vertebrates whose genomes have been annotated.
\end{itemize}
I evaluated the number of matching contigs, and for the non-redundant data sets, the number of uniquely matching contigs. Distinguishing between contigs that match and contigs that match uniquely is important, as despite my clustering during assembly, many contigs in the assembly appear redundant. These highly similar contigs likely result from misassemblies, allelic variants, alternative splicing isoforms, or recently duplicated paralogs. Parsing these categories is challenging without a reference genome and when expected coverage across contigs is uneven. Especially for projects interested in functional genomics, annotation of redundant contigs remains an important and unresolved issue. Here, I try to mitigate these errors by using reciprocal BLAST best matching to annotate contigs and selecting the best match. In doing so, I likely failed to annotate recently evolved paralogs, but I should not have multiple copies of the same gene in my downstream analyses.

Once I determined the best database both with respect to efficacy and efficiency, I used a custom script to annotate the contigs using a reciprocal best-match strategy via {\sffamily BLASTx} and {\sffamily tBLASTx}  \citep{Altschul1997} and defined the untranslated regions and coding sequence of the transcript using {\sffamily Exonerate} \citep{Slater2005}. Further, initial tests of the annotation pipeline uncovered two challenges: first, many contigs were chimeric and consisted of multiple, combined transcripts, and second, many of the predicted open reading frames (ORFs) had nonsense mutations, largely due to frameshift mutations.  To address these problems, I identified chimeric contigs using {\sffamily BLASTx} and split these contigs into individual genes, and I used the program {\sffamily FrameDP} to identify and correct for frameshift mutations \citep{Gouzy2009}. 

Finally, I searched unannotated contigs against the {\sffamily NCBI 'nr'} database using {\sffamily BLASTn} to determine these contigs' identity. As described in the {\sffamily Results}, these unannotated contigs largely went unidentified.  Thus, although some of these unannotated transcripts have viable open reading frames and/or had homologs in other lineages, and therefore, might be genes, I will be conservative and only use annotated transcripts in all downstream analyses. 

Finally, to describe the putative biological functions of my annotated contigs, I determined gene ontology using {\sffamily Blast2Go} \citep{Conesa2005}.

\subsection{Alignment}
The first step in identifying variants or estimating gene expression levels is to align the sequencing reads to one's reference genome. Here, I use my annotated transcripts as a pseudo-reference genome \citep{Wiedmann2008}, thus aligning the reads used to generate the assembly to the assembly itself. Here, I tested six different aligners ({\sffamily bowtie}, {\sffamily bowtie2}, {\sffamily bwa}, {\sffamily novoalign}, {\sffamily smalt}, {\sffamily SOAPaligner}, {\sffamily stampy}; \cite{Langmead2009,Langmead2012,Li2009,Lunter2011,Li2008}) to determine their efficacy and accuracy. These programs run the gamut of being fast but less sensitive to being slower and more sensitive. Here, sensitivity is defined as the aligner's ability to align reads with multiple mismatches. Previous results have shown \citep{Li2011} that alignment error is a common cause of miscalled SNPs, particularly alignment errors around indel sites. To evaluate these programs, I inferred genotypes from the alignments with {\sffamily SAMtools} \citep{samtools}. I then compared these genotypes to a small data set of known genotypes from one of the contact zones, \emph{C. rubrigularis} N/S. In another study, I had Sanger sequenced 200-400 bp of sequence from 10 to 15 genes for the same individuals sequenced here (Singhal, unpublished). Importantly, all these genes were represented at high coverage ($\ge$20$\times$) in this data set; thus, coverage is sufficiently great to ensure accurate genotype calling. I used these validated genotypes to determine the number of false positives and negatives in my inferred genotypes. Further, I evaluated these programs based on the proportion of reads and read pairs they aligned and the concordance of SNP calls across data sets.

\subsection{Variant discovery}
Two major types of variant discovery are SNP identification and genotype calling. Many researchers are interested only in identifying SNPs, or determining which nucleotide positions are variable in a sample of individuals. SNP-containing regions are then resequenced or genotyped for further analysis \citep{Wiedmann2008}. Increasingly, researchers are both identifying variable sites, and then, summarizing variation at these sites using the site frequency spectrum (SFS) or calling genotype likelihoods for each individual for subsequent population genomics analyses. SNP identification has become an easier exercise as sequencing costs dropped and coverage has increased. However, genotype calling remains a challenging proposition, particularly in diploid and polyploid individuals, as distinguishing heterozygosity, homozygosity, and sequencing errors at variable sites is difficult unless there is high coverage ($\ge$20$\times$, \cite{Nielsen2012}). Thus, I focus on genotype calling and its use in characterizing variation for population genomics analyses. Importantly, I assume in my approach and discussion that both alleles are expressed in each individual; although there are some data to suggest that expression can be allele-biased \citep{Palacios2009}, accounting for this complexity is beyond the scope of this study.

My results indicated that {\sffamily Bowtie2} was the most effective and efficient aligner (see \emph{Results}); thus, I used it for all downstream analyses. When identifying variants from alignment data, there are several approaches:
\begin{enumerate}
\item brute strength methods, in which the read counts for given alleles at a site are calculated, and variants are determined by an arbitrary cut-off (e.g. Yang et al 2011)
\item maximum likelihood ({\sffamily VarScan}) and Bayesian methods ({\sffamily GATK, SAMtools}) \citep{VarScan,GATK,samtools}, in which algorithms consider strand bias, alignment quality, base quality, and depth to call genotype likelihoods for individuals. These methods have been developed further to account for Hardy-Weinberg disequilibrium and linkage disequilibrium in calling and filtering variants \citep{samtools,GATK}, to use machine learning with a set of validated SNPs to improve algorithms \citep{GATK}, and to re-align reads near indel areas to ensure inaccurate alignments do not lead to false SNPs.
\item Bayesian methods ({\sffamily ANGSD}) which infer the site frequency spectrum for all the variants in the data set, which in in turn, is used a prior to estimate genotype likelihoods for individuals \citep{Nielsen2012}. This method is particularly useful for data sets with large population samples.
\end{enumerate}
Here, I test these three general types of SNP and genotype discovery, using read counting, {\sffamily  VarScan}, {\sffamily  samtools}, and {\sffamily ANGSD} in two sister lineage-pairs for which I have validated genotypes (\emph{C. rubrigularis} N/S and \emph{L. coggeri} N/C). I both looked at concordance of SNP and genotype calls across methods and calculated the number of false positives and negatives. 

\subsection{Homolog discovery}
Homologs between lineages must be identified for any comparative genomics analyses. In this study, my lineages are all closely-related, so homology identification is less challenging than in many other comparative studies. However, ensuring I am identifying orthologs across lineages and not paralogs is challenging, particularly as my annotation pipeline could not conclusively distinguish orthologs and paralogs in the absence of a reference genome. With that caveat, I test three different methods for identifying homology:
\begin{enumerate}
\item defining homologs by their annotation; \emph{i.e.}, contigs that share the same annotation are assumed to be homologs,
\item defining homologs by reciprocal best-hit BLAST, as is most commonly done in other studies \citep{MorenoHagelsieb2008},
\item the SNP method, or defining homologs by mapping reads from one lineage to the other lineages' assembly, identifying variants, and thus determining homologous sequence.
\end{enumerate}
I evaluated these methods by the number of homologs found, the percent of aligned sequence between homologs, and the raw number of differences between homologous sequence. I looked at homology discovery both between sister lineages and non-sister lineages, as I expect discovery across non-sister lineages will be harder.

\subsection{Biological inference}
Finally, I determined how robust biological inference is to the analysis method used. First, to determine how genotype calling affects downstream inference, I inferred the site frequency spectrum and associated summary statistics (Tajima's $D$, $\theta$, $\pi$) for one lineage across different genotype calling methods and different coverage levels. Second, to determine how homology identification affects downstream inference; I determined \emph{dN/dS} ratios and raw sequence divergence for each gene across different methods of homology.

\section{Results}
\subsection{Data Quality and Filtration}
Library preparation and sequencing were successful for all individuals. On average, I generated 3.5 $\pm$ 0.5 Gb per individual. Duplication rates, low-complexity sequences, and contamination levels were low (Table S2). However, aggressive filtering and merging significantly reduced the raw data set; I lost 27.1 $\pm$ 3.8\% of raw base pairs per individual. As seen in Figure S3, this strategy significantly improved the per-base quality of my data. Indeed, I was able to reduce sequencing error rates in my final data set five-fold (initial error rates: 0.3 $\pm$ 0.1\%, final error rates:  0.06 $\pm$ 0.01\%). These error rates are likely over-estimates, because I used the lower-quality portion of the read (the tail end) to identify sequencing errors. Despite this reduction in error rates, profiling of mismatches across the reads showed that both the head and tail of the read still harbor a higher number of mismatches compared to the rest of the read. This pattern persisted even when the first and last five base pairs of each read were trimmed prior to alignment (Fig. S4). Possibly, as others have found residual adaptor sequence in their data sets despite using rigorous adaptor trimming (Bi, unpublished; Almeida, unpublished), these heightened error rates could be due to adaptor sequences leading to misalignments and spurious SNPs. 

\subsection{\emph{de novo} Assembly}
To assemble my data, I tested five different programs, which employed different strategies (e.g., single k-mer, built-in multi-kmer approach, my custom multi-kmer approach). I evaluated the assemblies on many metrics; here, I show data for four of these metrics. With respect to the percentage of paired reads that aligned to the assembly, {\sffamily SOAPdenovo} and {\sffamily Trinity} performed far better than the rest of the assemblers (Fig. 2A), suggesting their assemblies were more contiguous. The same two assemblers and {\sffamily Velvet} also recovered the greatest number of annotated transcripts, measured here by the number of core eukaryotic genes found in these assemblies (CEGMA; \cite{Parra2007}; Fig. 2B).  {\sffamily OASES} and {\sffamily Trinity} appeared to be the most accurate, as they contained the fewest number of nonsense mutations in annotated ORFs (Fig. 2C). Finally, {\sffamily OASES}, {\sffamily Trinity} and {\sffamily SOAPdenovo} assemblies had the fewest number of putative chimeric transcripts (Fig. 2D). Looking across all these metrics, {\sffamily Trinity} emerges as the best assembler. Further, {\sffamily Trinity} did a good job assembling most of the data; on average, just 8.1 $\pm$ 4.3\% of contigs from other assemblies were unique to that assembly compared to {\sffamily Trinity}. As such, I used {\sffamily Trinity} assemblies for all downstream analyses. As seen in Table 1, the basic metrics of these assemblies (e.g., number of contigs, total length of assembly, and N50) were fairly constant across all lineages. Unlike other studies \citep{Comeault2012}, I find no correlation between contig length and coverage, suggesting my assembly is not data-limited (Fig. S5).

\subsection{Annotation}
After assembling the data, I annotated the assemblies in order to identify unique, annotated contigs for downstream analyses and to refine the assemblies further. First, because my focal lineages are evolutionarily distant from the nearest genome (MRCA $\approx$150 mya to \emph{Anolis carolinensis}), I wanted to test the efficacy of different databases to annotate my contigs. While more complete databases did lead more annotated contigs (Table S3), the increase was marginal. Further, larger databases consume significantly more computing time; here, annotating to the {\sffamily UniProt90} database took nearly 100 times the processor hours as annotating to \emph{A. carolinensis}. Thus, I used the \emph{A. carolinensis} database for all further annotations. Importantly, I could annotate these genomes to more distant relatives (\emph{G. gallus} and \emph{T. guttata}; MRCA $\approx$300 mya), without seeing a significant decrease in annotation success (Table S3). This result suggests such an annotation approach could work for organisms in even more genomically depauperate clades.

While annotating contigs, I identified a low percentage of chimeric contigs ($\approx$4\%), which I resolved by splitting these contigs into individual genes (Table S4). Inspecting alignments of sequencing reads to these chimeric contigs suggested that these contigs form during assembly and not due to technical errors during library preparation, as chimeric junctions generally had significantly reduced coverage. Further, a small portion of the predicted open reading frames (ORFs) of annotated contigs ($\approx$3\%) had premature stop codons. Although it is possible that these ORFs are pseudogenes \citep{KalyanaSundaram2012}, it seems more likely that they are due to assembly errors, as these contigs were generally highly expressed. Using {\sffamily FrameDP}, I was able to identify and fix many of these likely frameshift errors (Table S4).

Through this pipeline, I annotated an average of 23360 contigs per lineage, of which, an average of 11366 contigs matched to a unique gene (Table 1). I also recovered the full coding sequence for many genes; 67\% of unique annotated contigs encompassed the entire coding sequence for a gene, including portions of the 5' and 3' UTRs. These numbers appear reasonable -- the annotation for the \emph{A. carolinensis} genome currently includes 19K proteins, and liver tissue does not express all genes at a sufficiently high level to be represented here \citep{Ramskold2009}. These genes contribute to a diversity of biological processes and serve a wide range of molecular functions, suggesting I assayed a varied portion of the transcriptome (Fig. S6). 

Further, my pipeline appears to be robust; almost all unannotated contigs failed to find a good match in the NCBI {\sffamily 'nr'} database (Fig. S7). Approximately 9\% of unannotated contigs matched to genes; however, further analysis of these matches showed that almost all of them matched with such low-quality to prevent annotation. 

Additionally, by annotating contigs rigorously to limit the number of putative duplicate contigs, I significantly reduced the redundancy of my data set. When I aligned sequencing reads to my initial, unannotated assembly, I found that $\approx$10\% of mapped reads aligned to multiple places in the assembly, suggesting a high level of redundancy. After annotating the genome and removing redundant contigs, I reduced the percentage of mapped reads aligning non-uniquely to $\approx$2\%.  However, removing redundant contigs also lead to an average 21\% decline in reads mapped. Thus, it seems likely these redundant contigs are "biologically real", but we do not yet have the tools to parse such contigs properly \citep{Vijay2012}.

\subsection{Alignment}
Identifying variants and quantifying gene expression first require that sequencing reads are aligned to the reference genome. Here, I tested the efficacy of seven different alignment programs, which employ different algorithms over a range of sensitivity and speed. I evaluated these programs in three ways. First, I used my externally validated set of genotypes to see how many genotypes were inferred correctly. Almost all of the aligners performed well and led to the correct genotype at $\ge$90\% of the sites. Although the false negative rate was moderately high ($\approx$5\% for most aligners), the false positive rate was low (Table 2). {\sffamily Bowtie2} clearly outperformed the rest of the aligners and was thus used for all downstream analyses. Second, I evaluated how many read pairs and reads the programs could align. Although {\sffamily Novoalign}, {\sffamily smalt} and {\sffamily stampy} are generally considered to be more sensitive aligners, I found little variation in the percentage of reads aligned across programs (Fig. 3). {\sffamily Bowtie2} and {\sffamily stampy} were able to align the most paired reads, which is useful as aligning paired reads reduces the likelihood of errant matches and non-unique matches \citep{Bao2011}. Finally, I looked at overlap in SNPs inferred across programs. Problematically, although all programs were fed the same reference genome and sequencing reads, I saw only moderate overlap -- on average, only 77$\pm$9\% of SNPs were shared. Checking the raw alignments suggested these discrepancies often arose from differences in alignment rather than differences in SNP inference post-alignment. These results suggest that alignment is likely a major source of error in \emph{de novo} HTS analyses, as has been suggested by other studies \citep{Li2011,Lin2012,Kleinman2012}. That said, when the same SNPs were called across programs, genotype inference was highly concordant; 94$\pm$2\% of genotype calls were the same across alignment methods, and inferred allele frequency at these SNPs was highly correlated ($r$=0.94$\pm$0.01). 

\subsection{Variant Discovery}
After alignment, programs for variant inference are used to call SNPs and genotypes. In the previous tests, I used the variant discovery program {\sffamily SAMtools} for all analyses; here, I test a few approaches: a brute strength approach, in which I call SNPs and genotypes based solely on count data, two probabilistic methods ({\sffamily SAMtools} and {\sffamily VarScan}), and a probabilistic method that uses the allele frequency spectrum ({\sffamily ANGSD}). I first assessed accuracy of genotype calls by using my externally validated genotype set. In general, I found that all methods performed fairly well -- particularly, when a SNP was identified, all programs inferred the correct genotype with high accuracy ($\ge$98\%; Table 3). However, the count method of identifying variation led to many false positives, an unsurprising result given its failure to account for sequence error or alignment score. {\sffamily ANGSD} had a high false negative rate, the reason for which is unclear, though is possibly due to the small sample sizes used here. But, as shown by other work, {\sffamily ANGSD} is best suited for correctly inferring the shape of the site frequency spectrum \citep{Nielsen2012}.  Comparing across all SNPs found across all programs, I found that concordance across all SNPs was moderate, similar to my comparative alignment results. On average, only 83\% of SNP calls are shared across programs. More promisingly, when a site is inferred as a SNP, 98\% of the genotype calls are shared across programs. Overall, these results suggested {\sffamily SAMtools} performed the best, so I used it for all downstream analyses.

Upon defining SNPs and then genotypes for each individual, I explored how different variant discovery methods affect biological inference by constructing the SFS. Despite the only moderate levels of concordance in SNP calls, I find that the SFS is nearly identical across all the different approaches but {\sffamily VarScan} (Fig. 4). Importantly, this result only holds true when I restrict analysis to higher-coverage contigs ($\ge$10$\times$); low-coverage contigs show aberrant patterns. Although the SFS is similar across all approaches, estimates of key population genetic summary statistics (\emph{i.e.}, $\theta_{w}$, $\pi$) vary depending on the approach -- an unsurprising result given that the total number of SNPs inferred differs across approaches. Thus, prior to using these data for population genetic analyses, ascertainment bias must be factored into any downstream inference (citation). Finally, to look at these SNPs in greater detail, I annotated the SNPs I found in two sister-lineages, with respect to how they are segregating, their location relative to the gene, and their coding type (Fig. S9). Not only are the patterns of polymorphism and non-synonymous/synonymous mutations reasonable \citep{Begun2007}, but I also see that I have many types of variants (\emph{i.e.}, coding vs. non-coding, non-synonymous vs. synonymous, fixed vs. polymorphic), which will permit me to use the data to look for adaptive signatures of molecular evolution, infer demographic history, and to develop markers.

\subsection{Homolog discovery}
To identify homologs between lineages, I tested three different methods and then evaluated their effectiveness. All three methods performed well, identifying more than 8000 homologous pairs between lineages within-genera and between-genera for a significant portion of the contig length (Fig. 5). However, with the SNP method for homology, alignment efficiency dropped off significantly in between-genera comparisions, leading to identified homologs being shorter. I chose to use reciprocal BLAST matching to identify homologs for all downstream analyses as it was able to identify more homologs than the two other methods and it worked well across evolutionary distances (Fig. 5). This approach identified 8800 homologous contigs across all seven lineages for use in comparative analyses. 

Estimation of the summary statistics (sequence divergence  and $dN/dS$ ratios between homologs from lineage-pairs) is affected by how homologs are defined (Fig. S10). Defining homologs via annotation or via reciprocal BLAST matching gives very similar results for both sequence divergence and $dN/dS$. However, using SNPs to reconstruct the homolog results in a fuzzier pattern. This discrepancy likely stems from the many homologs for which coverage is low ($<$10$\times$), and thus, SNP inference is error-ridden (see \emph{Results: Variant Discovery}). Thus, this method for homolog identification should account for differences in coverage, where appropriate. 

\section{Discussion}
In creating and implementing a pipeline for high-throughput sequence data, I noted several possible sources of error (Fig. 1B):
\begin{enumerate}
\item Errors introduced during library preparation, which can include human contamination, errors introduced during PCR amplification of the library, and cross-contamination between samples
\item Errors introduced during sequencing, the frequency and type of which are dependent on the chemistry of sequencing platform, and subsequent de-multiplexing
\item Errors introduced during assembly \citep{Baker2012}, such as misassembly of reads to create chimeric contigs
\item Errors due to misalignment of reads to assembly during variant discovery, particularly caused by indels in alignments and reads that map to multiple locations
\item Errors in SNP and genotype calling, such as not sampling both alleles and thus mistakenly calling a homozygote
\end{enumerate}
To this, I add two additional sources of uncertainty that every study in evolutionary genomics faces -- have contigs been annotated correctly and have orthologs between compared genomes been identified correctly \citep{Chen2007}? Errors can arise at any stage in the process; such errors percolate through subsequent steps, likely affecting all downstream inference \citep{Vijay2012,Lin2012,Kleinman2012}. Whether using their own pipeline or a pre-existing pipeline, researchers will want to incorporate some of the checks suggested here to ensure that the pipeline is working well for their data and that incidence of errors is low. Moving forward, the questions become how to limit these errors and how to mitigate their effects. 

All these sources of error are non-trivial, but with careful data checking and willingness to discard low-quality data, I could mitigate the effects of these errors. I address these sources of error by stepping through the pipeline, explaining how I was able to reduce the error and identifying areas of future work. First, as has now become standard, scrubbing reads for low-quality bases and adaptors is a must -- as shown here, read cleaning can reduce error rates noticeably. When possible, merging reads from paired-end reads can further decrease error rates and will lead to more accurate estimates of coverage for expression studies \citep{Magoc2011}. Second, having a high-quality assembly is crucial both for accurate annotation and variant discovery. Inferring the quality of \emph{de novo} assemblies is challenging, as there are no clear metrics or comparisons to use \citep{Martin2011}. However, I propose a few metrics, which can be used with transcriptome data -- primarily, looking for assemblies that minimize chimerism and non-sense mutations, that are contiguous, and that capture a significant portion of known key genes. Undoubtably, errors remain in the final assemblies, but these metrics helped me select the most accurate assembly for downstream analyses. Additionally, contig redundancy in final assemblies remains a pressing challenge. By using a strict reciprocal-BLAST annotation strategy, I removed many of these apparently redundant contigs. However, this approach certainly removed some biologically real contigs that were recent duplicates and alternative splicing isoforms fo interest to those interested in expression differences between biological groups \citep{Vijay2012}. Researchers should continue to explore better methods to identify orthologs and paralogs.

Alignment and variant discovery remain notable challenges. In part, a poor-quality assembly genome truly can affect variant discovery -- alignments across misassemblies can led to errant SNP calls, particularly when misassemblies introduce indels \citep{Li2011}. Further, unless some sort of redundancy reduction is used, many contigs will be nearly identical in an assembly, leading to a high rate of non-unique alignments and miscalled SNPs. I was able to remove most redundant contigs, and thus, I reduced the proportion of non-unique alignments. I still see evidence for errors in alignment as (1) discrepancies between our externally-validated SNP set and genotype calls from these alignments and (2) the only moderate level of congruence between different approaches fueled by the same data. The same patterns hold for genotype inference based on alignment. My work here suggests, that given the data I have, the best approach is to rely on contigs with higher coverage -- 10 to 20$\times$, at least -- and to account for this ascertainment bias in any biological inference. 

Further, to ensure the vagaries of variant discovery do not unduly influence our biological inference, we should use the genotype likelihoods and not genotype calls for downstream work. Ideally, researchers would conduct subsequent inference that use the SFS or genotype likelihoods as input, such as {\sffamily BAMOVA} \citep{Gompert2011} or {\sffamily dadi} \citep{Gutenkunst2009}, thus ensuring uncertainty in SNP and genotype calling is incorporated into model fitting. However, many analyses, particularly those used by most biodiversity researchers (\emph{i.e.}, coalescent-based demography and phylogeny programs), require known genotypes or haplotypes. Until uncertainty is incorporated into such programs, researchers will have to arbitrarily chose cutoffs to determine most likely genotypes. In such cases, researchers might want to restrict their analyses to regions with high coverage, where calls are likely more certain \citep{Nielsen2012}. 

Moving forward, how can we reduce the sources of errors stemming from alignment errors and genotype inference? Improved assemblies, facilitated by new long-read sequencing technologies, will certainly help. As researchers collect externally validated SNP data sets, they can use programs like {\sffamily GATK} to recalibrate variant calling and to realign around indels \citep{GATK}.  Researchers will also increasingly sequence more individuals in a population, which will better take advantage of multi-sample methods like {\sffamily samtools} and {\sffamily ANGSD} \citep{samtools,Nielsen2012}. Finally, programs like {\sffamily Cortex}, which assemble across individuals to provide both a reference assembly and individual assemblies, are promising. Simulations suggest that this method can also better handle data with indel polymorphism \citep{Iqbal2012}. 

Finally, homolog discovery is a challenge in any genome project \citep{Chen2007}, and this project was no exception. All three methods I tested for homolog discovery worked well, but I recommend only using a SNP-based approach between lineages that are closely-related and for contigs with high coverage. Moving forward, as we acquire more comparative genomic data across the tree of life, homolog discovery should become an easier problem, as fueled by comparative clustering programs like {\sffamily OrthoMCL} \citep{Chen2005}.

Through this work, I collated a large data set of over 12K annotated contigs, spanning a wide-range of biological functions, and over 100K SNPs between lineage-pairs, spanning a wide-range of locations and coding types. Notably, I was able to do all of these analyses using existing, open-source software and, but for assembly, by using a low-end desktop machine. Genomic analyses are not just for those working with humans or mice anymore. With careful and thoughtful data curation, HTS can enable researchers to use genomic approaches to explore all the branches in the tree of life.

\section{Acknowledgements}
I gratefully acknowledge M. Chung, J. Penalba, and L. Smith for technical support, and the Seqanswers.com community for providing timely and thoughtful advice. R. Bell, K. Bi, J. Bragg, T. Linderoth, M. MacManes, C. Moritz, R. Nielsen, S. Ramirez, F. Zapata, and members of the Moritz Lab Group provided comments and suggestions during this work and on this manuscript that greatly improved its quality. Financial support for this work was provided by National Science Foundation (Graduate Research Fellowship and Doctoral Dissertation Improvement Grant), the Museum of Vertebrate Zoology Wolff Fund, and a Rosemary Grant Award from the Society of the Study of Evolution. This work was made possible by the supercomputing resources provided by NSF XSEDE, in particular the clusters at Texas Advanced Computing Center and Pittsburgh Supercomputing Center.

\clearpage

\bibliographystyle{evolution}
 \bibliography{bibliographyGenomics}
 
 \clearpage

\section{Tables}

\begin{table}[H]
\centering
\label{graph}
\rowcolors{2}{gray!20}{}
\begin{tabular}{ >{\centering\arraybackslash}m{3cm} >{\centering\arraybackslash}m{1.5cm} >{\centering\arraybackslash}m{1.5cm} >{\centering\arraybackslash}m{1.5cm} >{\centering\arraybackslash}m{1.5cm} >{\centering\arraybackslash}m{1.5cm} >{\centering\arraybackslash}m{2cm}  }
\rowcolor[gray]{0.63} assembly & number contigs & total length & n50  & annotated contigs & annotated contigs (unique) & complete annotated contigs \\ 
\emph{C. rubrigularis}, N & 104648 & 89.1e6 & 1806  & 25198 & 12063 & 8179 \\
\emph{C. rubrigularis}, S & 98280 & 84.3e6 & 1780   & 24323 & 11558 & 7697 \\
\emph{L. coggeri}, N & 96798 & 87.5e6 & 1972   & 22760 & 11457 & 7344   \\
\emph{L. coggeri}, C & 106937 & 92.7e6 & 1845  & 23852 & 10894 & 7796 \\
\emph{L. coggeri}, S & 112935 & 89.6e6 & 1549  & 23774 & 11029 & 7258  \\
\emph{S. basiliscus}, C & 84756 & 77.7e6 & 1951  & 21584 & 11221 & 7586 \\
\emph{S. basiliscus}, S & 98685 & 83.5e6 & 1749  & 22031 & 11340 & 7696 \\
\end{tabular}
\caption{Summary of assemblies and their annotation. Complete annotated contigs are those with some 5' and 3' UTR sequence, as well as the full coding sequence.}
\end{table}

\begin{table}[H]
\centering
\label{graph}
\rowcolors{2}{gray!20}{}
\begin{tabular}{>{\centering\arraybackslash}m{3cm} >{\centering\arraybackslash}m{1.5cm} >{\centering\arraybackslash}m{1.5cm} >{\centering\arraybackslash}m{1.5cm} >{\centering\arraybackslash}m{1.5cm} >{\centering\arraybackslash}m{1.5cm} >{\centering\arraybackslash}m{1.5cm} >{\centering\arraybackslash}m{1.5cm}}
\rowcolor[gray]{0.63} genotype & {\sffamily bowtie} &  {\sffamily bowtie2} &  {\sffamily bwa} &  {\sffamily novoalign} &  {\sffamily smalt} &  {\sffamily SOAPaligner} &  {\sffamily stampy} \\ 
right genotype & 379 (89.8\%) &	419 (99.2\%) & 	381 (90.3\%) & 383 (90.8\%) & 393 (93.1\%) & 207 (49.0\%) & 391 (92.7\%) \\
wrong genotype & 29	(6.9\%) & 3 (0.7\%) &	7 (1.7\%) & 9 (2.1\%) & 6	(1.4\%) & 52 (12.3\%) & 8 (1.9\%) \\
false negative & 12	(2.8\%) & 0 (0\%) & 34 (8.1\%) & 30 (7.1\%) & 23 (5.5\%) & 163 (38.6\%) & 23 (5.5\%) \\
false positive & 3	& 1	 & 1	& 1	& 1 & 1 & 5 \\
\end{tabular}
\caption{Accuracy of genotype inference following the use of different programs for alignment; all genotypes were inferred using {\sffamily samtools} post-alignment. Parenthetical percentages show the relative proportions of genotype types.}
\end{table}

\begin{table}[H]
\centering
\label{graph}
\rowcolors{2}{gray!20}{}
\begin{tabular}{>{\centering\arraybackslash}m{3cm} >{\centering\arraybackslash}m{3cm} >{\centering\arraybackslash}m{3cm} >{\centering\arraybackslash}m{3cm} >{\centering\arraybackslash}m{3cm}}
\rowcolor[gray]{0.63} Genotype & {\sffamily ANGSD} & count data & {\sffamily SAMtools} & {\sffamily VarScan} \\ 
right genotype & 520 (68.4\%) & 745 (98.0\%) & 750 (98.7\%) & 745 (98.0\%) \\
wrong genotype & 3 (0.3\%) & 15 (2.0\%) & 10 (1.3\%) & 15 (2.0\%) \\
false negative & 230 (30.2\%) & 0 (0\%) & 0 (0\%) & 0 (0\%) \\
false positive & 6 & 134  & 1  & 12  \\
\end{tabular}
\caption{Accuracy of genotype inference across different programs for genotype inference; for all, {\sffamily Bowtie2} was used for alignment. Parenthetical percentages show the relative proportions of genotype types.}
\end{table}

\section{Figures}

\begin{figure}[H]
\centering	  
\includegraphics{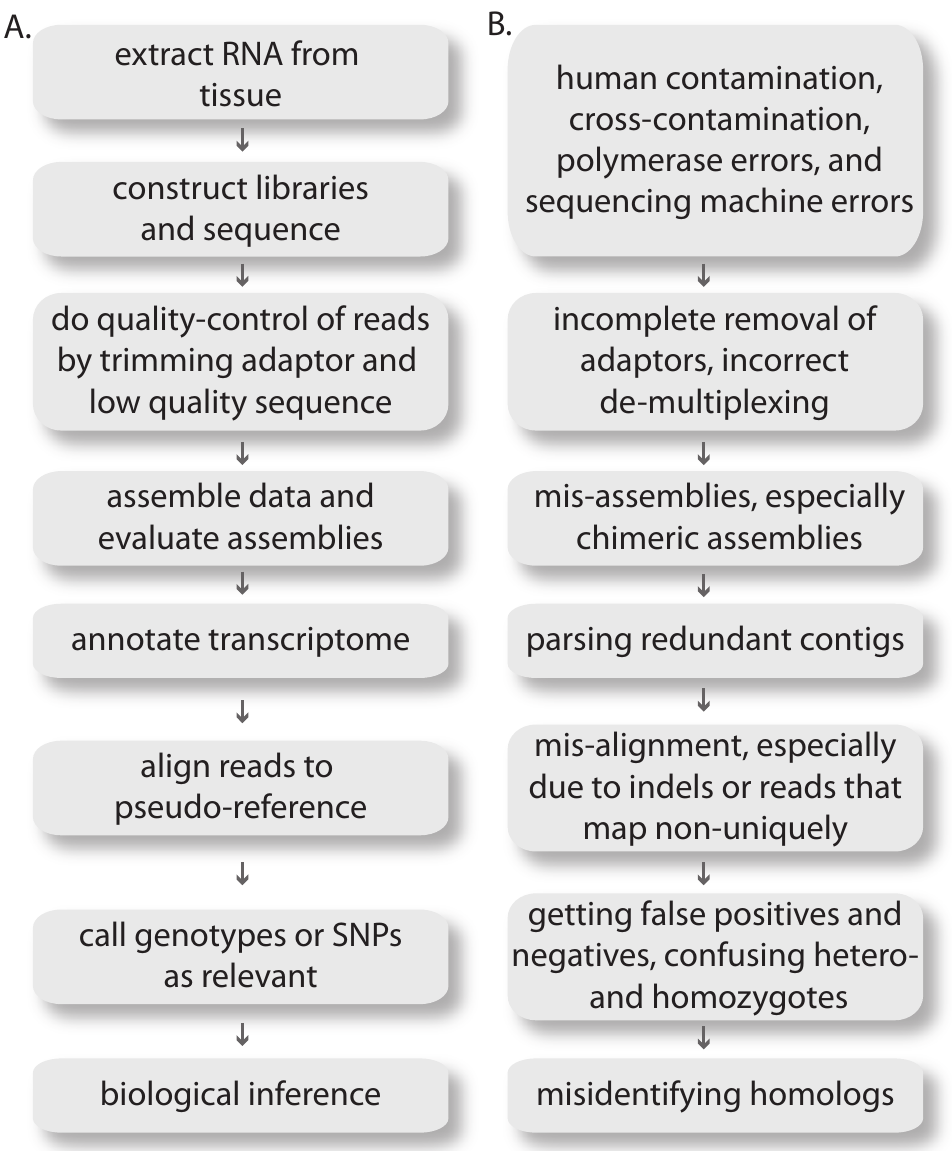}
	\caption{A. Pipeline for handling transcriptome data for \emph{de novo} population genomic analyses, as presented in this study. B. Errors introduced at each stage in the pipeline.}
\end{figure}

\begin{figure}[H]
\centering	
\includegraphics{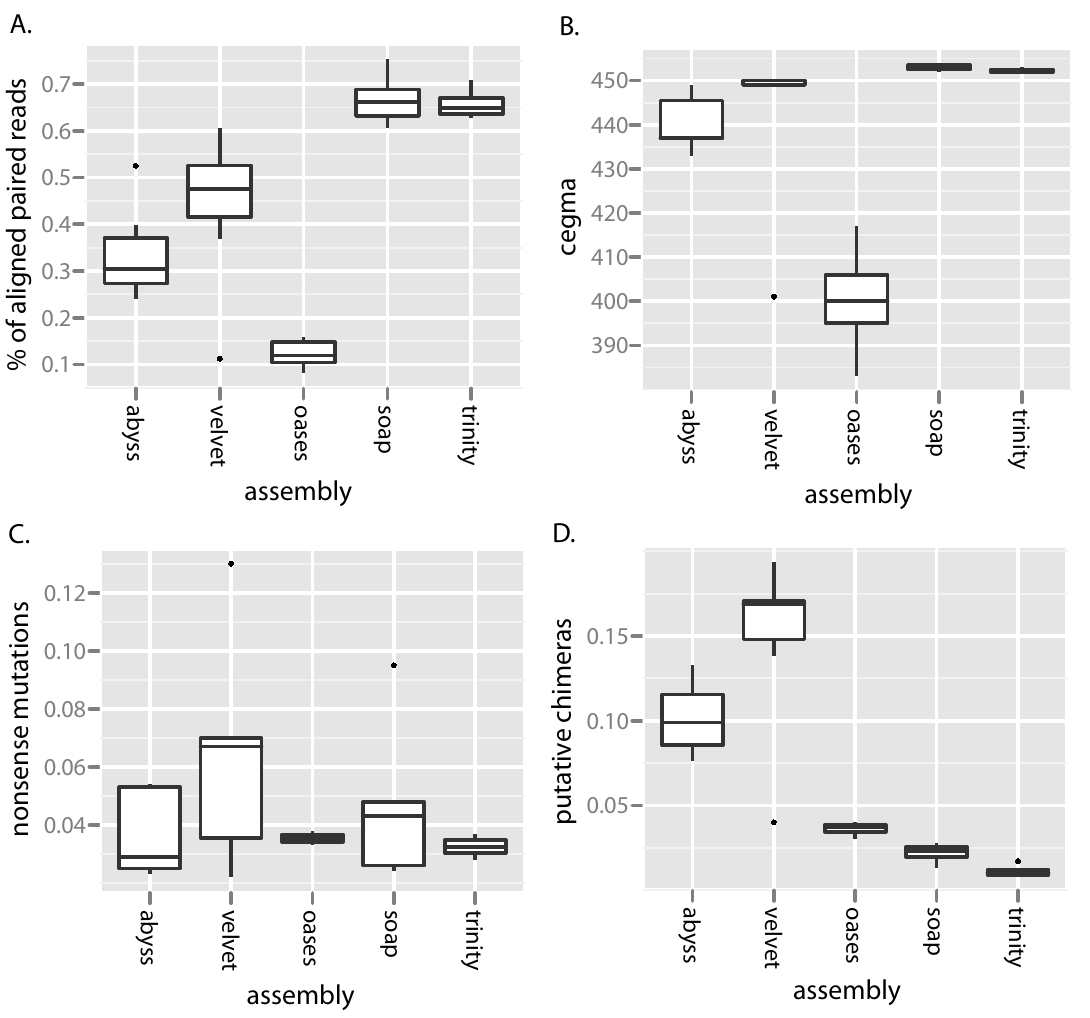}
	\caption{Evaluation of assemblies across the seven sequenced lineages according to A. percentage of paired reads that aligned to reference, B. number of CEGMA genes that are found in assembly, C. percentage of annotated coding sequences that had nonsense mutations, and D. percentage of contigs that were putative chimeras.}
\end{figure}

\begin{figure}[H]
\centering	
\includegraphics{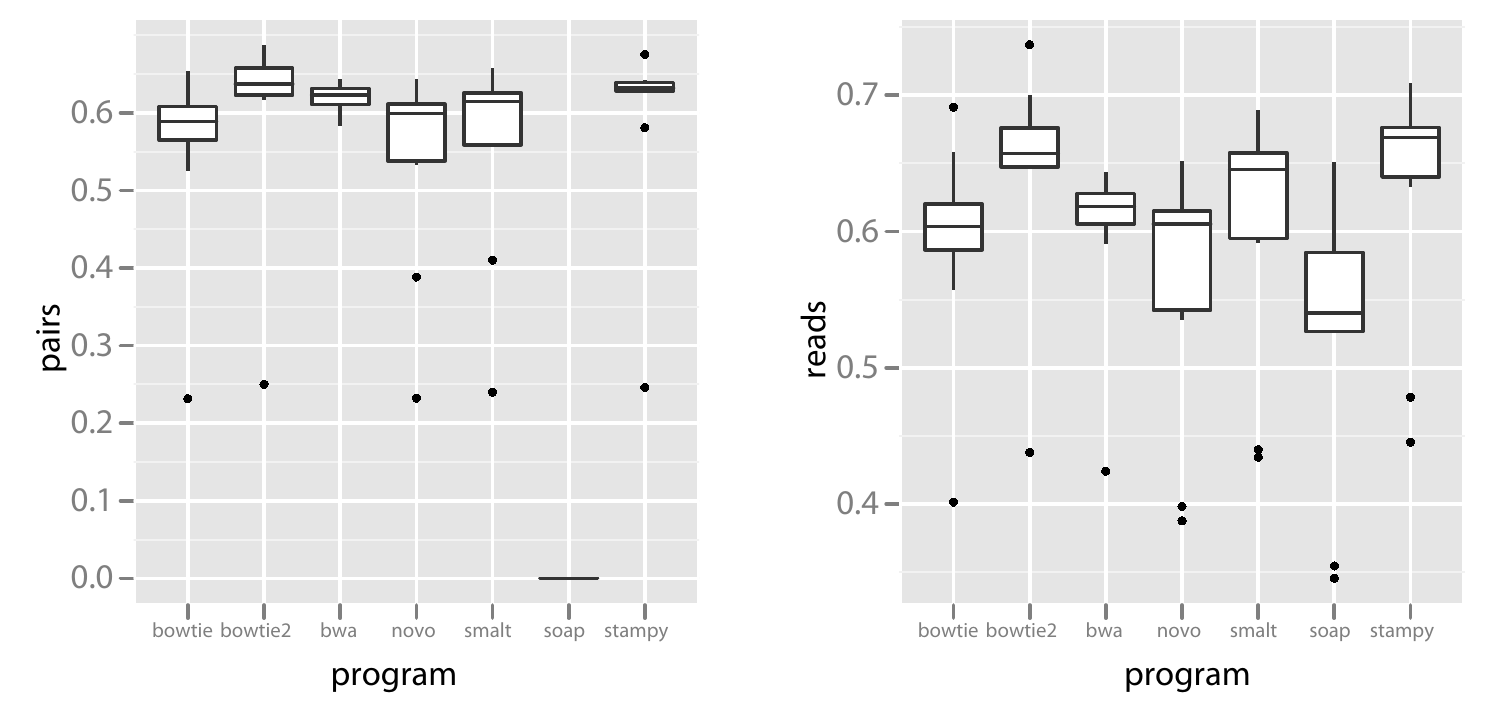}
	\caption{Evaluation of different alignment software across three randomly selected lineages with respect to two metrics, A. number of paired reads aligned and B. number of reads aligned.}
\end{figure}

\begin{figure}[H]
\centering	  
\includegraphics[width=\textwidth]{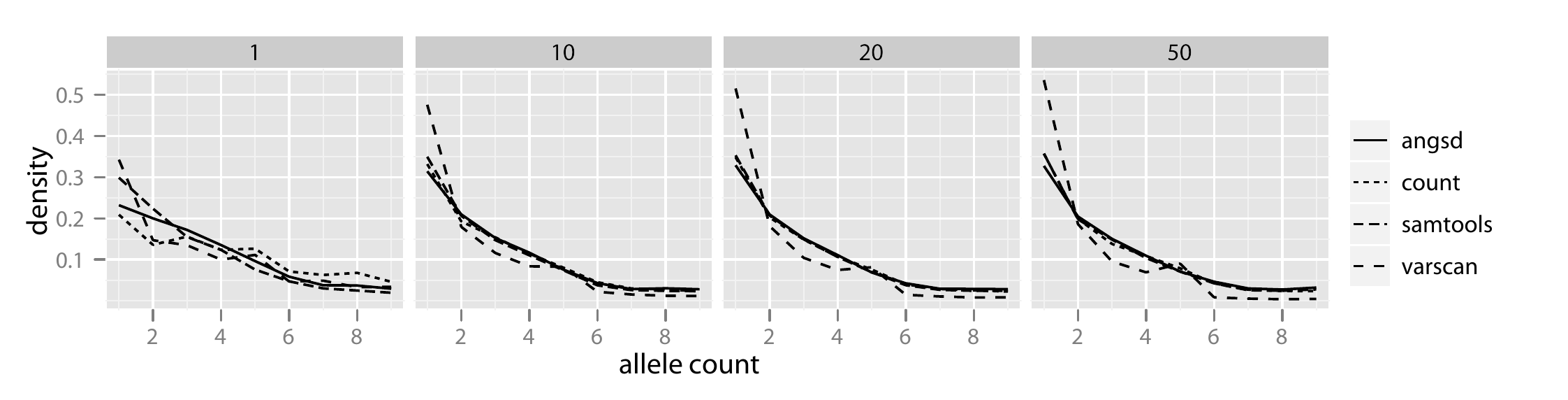}
	\caption{Unfolded allele frequency spectrum for variants within a randomly selected lineage for sites represented at 1{\sffamily x}, 10{\sffamily x}, 20{\sffamily x}, and 50{\sffamily x} coverage per individual, across different methods for genotype inference.}
\end{figure}

\begin{figure}[H]
\centering	 
\includegraphics{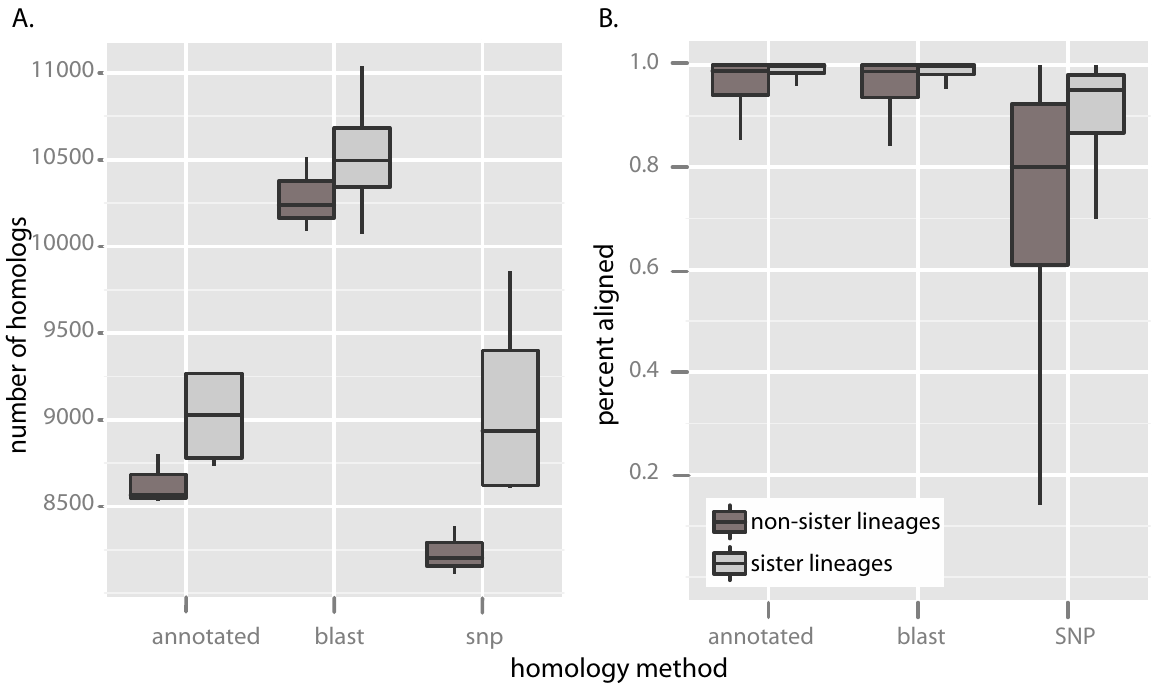}
	\caption{Summary of different methods for homolog discovery between all lineage comparisons of interest, considering A. number of homologs for which 75\% of sequence was aligned and B. percent of homolog aligned.}
\end{figure}

\setcounter{figure}{0}
\setcounter{table}{0}

\section{Supplementary Tables}

\begin{table}[H]
\centering
\label{graph}
\rowcolors{2}{gray!20}{}
\begin{tabular}{ >{\centering\arraybackslash}m{1.7 cm} >{\centering\arraybackslash}m{3.2 cm} >{\centering\arraybackslash}m{1.5 cm} >{\centering\arraybackslash}m{1.5 cm} >{\centering\arraybackslash}m{6 cm} }
\rowcolor[gray]{0.63} individual & lineage & latitude & longitude & Locality \\ 
SS34 & \emph{C. rubrigularis} N & -16.617 & 145.458 & Mount Harris \\
SS35 & \emph{C. rubrigularis} N & -16.617 & 145.458 & Mount Harris \\
SS37 & \emph{C. rubrigularis} N & -16.611 & 145.452 & Mount Harris \\
SS40 & \emph{C. rubrigularis} N & -16.611 & 145.452 & Mount Harris \\
SS41 & \emph{C. rubrigularis} N & -16.611 & 145.452 & Mount Harris \\
SS48 & \emph{C. rubrigularis} S & -17.694 & 145.694 & S. Johnstone River, Sutties Gap Rd \\
SS50 & \emph{C. rubrigularis} S & -17.694 & 145.694 & S. Johnstone River, Sutties Gap Rd \\
SS52 & \emph{C. rubrigularis} S & -17.660 & 145.722 & S. Johnstone River, Sutties Gap Rd \\
SS56 & \emph{C. rubrigularis} S & -17.678 & 145.710 & S. Johnstone River, Sutties Gap Rd \\
SS57 & \emph{C. rubrigularis} S & -17.678 & 145.710 & S. Johnstone River, Sutties Gap Rd \\
SEW08448 & \emph{L. coggeri} C & -16.976 & 145.777 & Lake Morris Rd \\
SEW08452 & \emph{L. coggeri} C & -16.976 & 145.777 & Lake Morris Rd \\
SS135 & \emph{L. coggeri} C & -16.976 & 145.777 & Lake Morris Rd \\
SS136 & \emph{L. coggeri} C & -16.976 & 145.777 & Lake Morris Rd \\
SS138 & \emph{L. coggeri} C & -16.976 & 145.777 & Lake Morris Rd \\
SS64 & \emph{L. coggeri} N & -16.579 & 145.315 & Mount Lewis \\
SS65 & \emph{L. coggeri} N & -16.572 & 145.322 & Mount Lewis \\
SS67 & \emph{L. coggeri} N & -16.578 & 145.308 & Mount Lewis \\
SS72 & \emph{L. coggeri} N & -16.585 & 145.289 & Mount Lewis \\
SS74 & \emph{L. coggeri} N & -16.584 & 145.302 & Mount Lewis \\
SS54 & \emph{L. coggeri} S & -17.660 & 145.722 & S. Johnstone River, Sutties Gap Rd \\
SS59 & \emph{L. coggeri} S & -17.700 & 145.693 & S. Johnstone River, Sutties Gap Rd \\
SS60 & \emph{L. coggeri} S & -17.700 & 145.693 & S. Johnstone River, Sutties Gap Rd \\
SS62 & \emph{L. coggeri} S & -17.676 & 145.713 & S. Johnstone River, Sutties Gap Rd \\
SS63 & \emph{L. coggeri} S & -17.628 & 145.740 & S. Johnstone River, Sutties Gap Rd \\
SS25 & \emph{S. basiliscus} C & -17.295 & 145.712 & Butchers Creek \\
SS28 & \emph{S. basiliscus} C & -17.299 & 145.701 & Butchers Creek \\
SS29 & \emph{S. basiliscus} C & -17.299 & 145.701 & Butchers Creek \\
SS30 & \emph{S. basiliscus} C & -17.299 & 145.701 & Butchers Creek \\
SS32 & \emph{S. basiliscus} C & -17.299 & 145.701 & Butchers Creek \\
SS127 & \emph{S. basiliscus} S & -18.199 & 145.849 & Kirrama Range Rd \\
SS128 & \emph{S. basiliscus} S & -18.199 & 145.849 & Kirrama Range Rd \\
SS129 & \emph{S. basiliscus} S & -18.199 & 145.849 & Kirrama Range Rd \\
SS130 & \emph{S. basiliscus} S & -18.199 & 145.849 & Kirrama Range Rd \\
SS131 & \emph{S. basiliscus} S & -18.199 & 145.849 & Kirrama Range Rd \\
\end{tabular}
\caption{Individuals included in this study and their associated locality data.}
\end{table}

\begin{table}[H]
\centering
\label{graph}
\rowcolors{2}{gray!20}{}
\begin{tabular}{ >{\centering\arraybackslash}m{4cm} >{\centering\arraybackslash}m{4cm} }
\rowcolor[gray]{0.63} filtering type & rate \\
duplication & 1.4 $\pm$ 0.2\% \\
contamination & 0.4 $\pm$ 1.1\% \\
low-complexity reads & 0.004 $\pm$ 0.003\% \\
merging reads & 68.7 $\pm$ 4.7\% \\
\end{tabular}
\caption{Quality control filtering and their rates in raw data, summarized across seven lineages.}
\end{table}

\begin{table}[H]
\centering
\label{graph}
\rowcolors{2}{gray!20}{}
\begin{tabular}{>{\centering\arraybackslash}m{5cm} >{\centering\arraybackslash}m{3cm} >{\centering\arraybackslash}m{3cm}}
\rowcolor[gray]{0.63} database & annotated contigs & unique, annotated contigs \\ 
\emph{A. carolinensis} & 23804 & 12218 \\
\emph{G. gallus} & 22324 & 11146 \\
{\sffamily UniProt90} database & 26089 & 12324 \\
{\sffamily Ensembl} 9-species database & 25838 & NA \\
{\sffamily Ensembl} 54-species database & 26601 & NA \\
\end{tabular}
\caption{Number of contigs annotated according to different reference databases for a randomly selected assembly.}
\end{table}

\begin{table}[H]
\centering
\label{graph}
\rowcolors{2}{gray!20}{}
\begin{tabular}{ccccc}
\rowcolor[gray]{0.63} assembly & initial chimerism & final chimerism & initial stop codons & final stop codons \\ 
\emph{C. rubrigularis}, N & 4.6\% & 0.0\% & 2.6\% & 0.6\% \\
\emph{C. rubrigularis}, S & 3.7\% & 0.0\% & 2.8\% & 0.8\% \\
\emph{L. coggeri}, N & 10.3\% & 0.0\% & 3.3\% & 1.1\% \\
\emph{L. coggeri}, C & 5.5\% & 0.0\% & 3.1\% & 1.0\% \\
\emph{L. coggeri}, S & 3.9\% & 0.0\% & 3.3\% & 1.0\% \\
\emph{S. basiliscus}, C & 4.4\% & 0.0\% & 2.6\% & 0.6\% \\
\emph{S. basiliscus}, S & 4.0\% & 0.0\% & 2.8\% & 0.7\% \\
\end{tabular}
\caption{Prevalence of chimerism, or percentage of contigs that appeared to consist of multiple genes misassembled together, and stop codons, or percentage of contigs that had nonsense mutations, in assemblies, summarized across seven lineages both before and after the data were run in the annotation pipeline.}
\end{table}

\begin{table}[H]
\centering
\label{graph}
\rowcolors{2}{gray!20}{}
\begin{tabular}{>{\centering\arraybackslash}m{3cm} >{\centering\arraybackslash}m{3cm} >{\centering\arraybackslash}m{3cm}}
\rowcolor[gray]{0.63} coverage & number of contigs within lineage & number of contigs between lineages \\ 
10{\sffamily x} & 3326 $\pm$ 494 & 2606 $\pm$ 399 \\
20{\sffamily x} & 1888 $\pm$ 316 & 1439 $\pm$ 245 \\
30{\sffamily x} & 1311 $\pm$ 245 & 981 $\pm$ 178 \\
40{\sffamily x} & 994 $\pm$ 190 & 741 $\pm$ 133 \\
50{\sffamily x} & 808 $\pm$ 157 & 602 $\pm$ 108 \\
\end{tabular}
\caption{Number of annotated contigs which have given coverage for each individual; shown for one randomly selected lineage-pair.}
\end{table}

\section{Supplementary Figures}

\begin{figure}[H]
\centering	  
\includegraphics{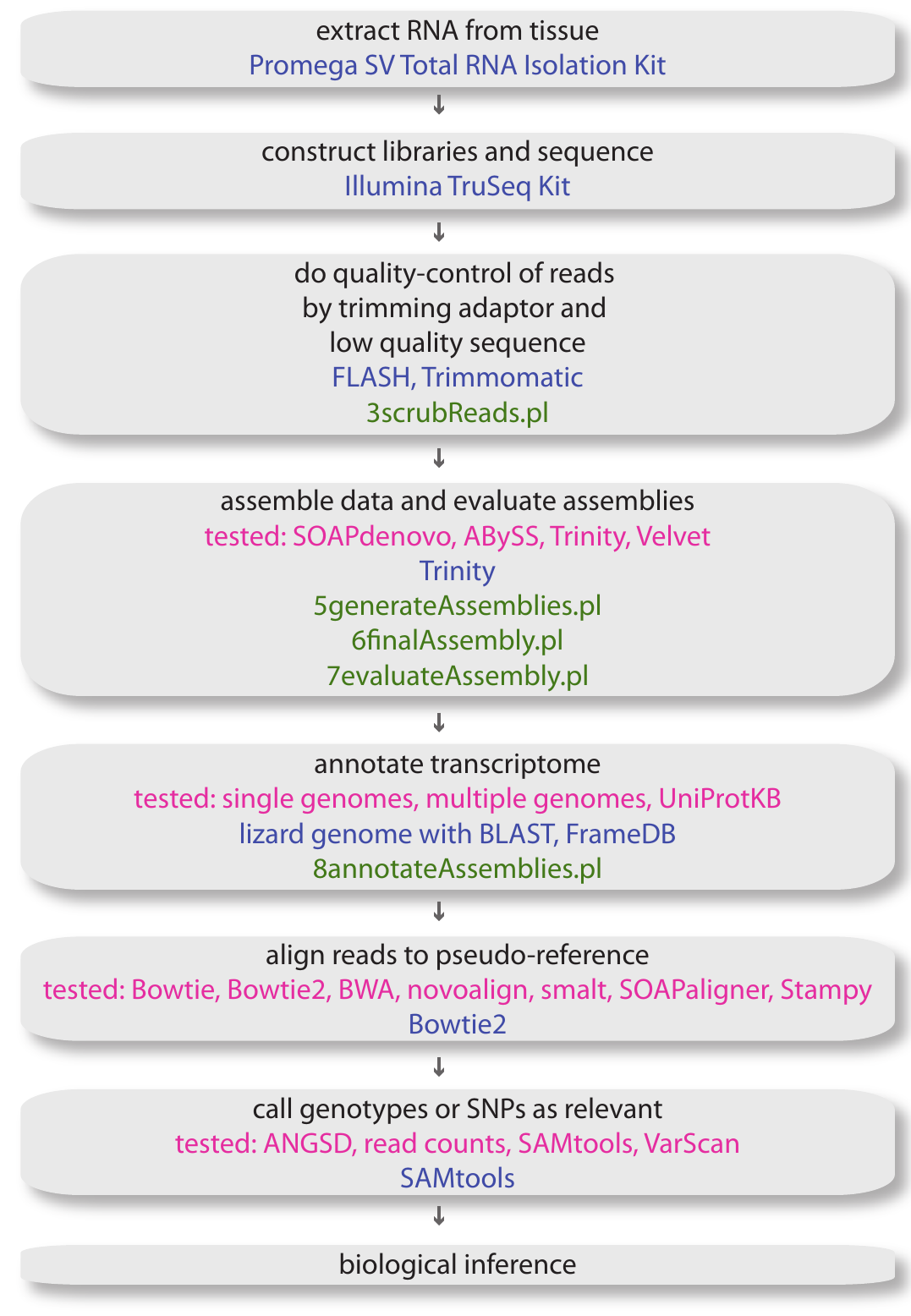}
	\caption{Pipeline used in this work, annotated to show (1) different approaches tested [pink], (2) the approach used for the final analysis [blue], and (3) scripts used, as named in the DataDryad package [green].}
\end{figure}

\begin{figure}[H]
\centering	  
\includegraphics[width=6in]{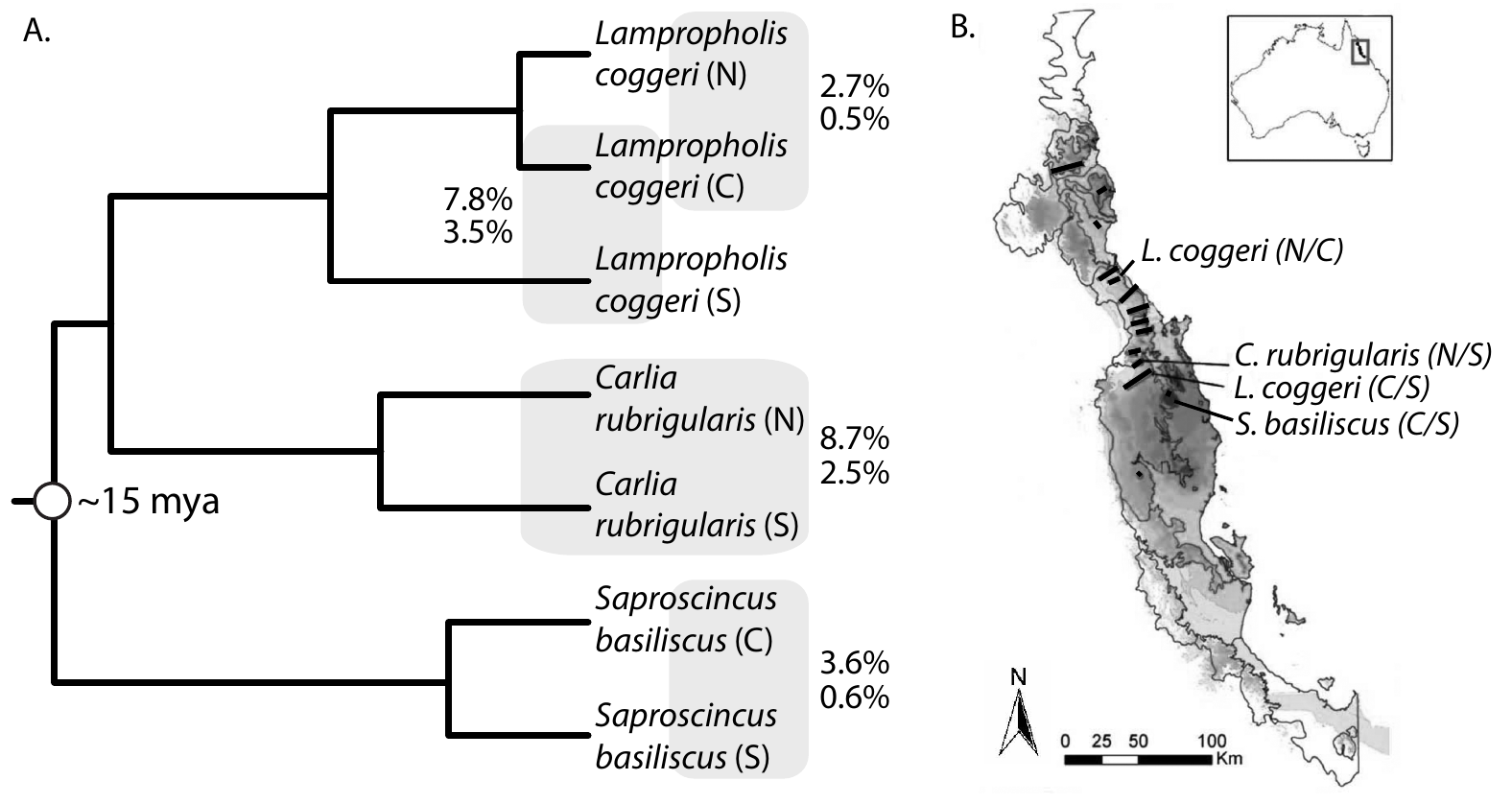}
	\caption{A. Phylogeny of the lineages studied in this work. Boxes indicate contacts studied; the top percentage reflects the mitochondrial divergence between lineages and the bottom is nuclear. B. A map of the Australian Wet Tropics, with all identified contact zones represented by black lines. Contacts of interest in this study are labelled.}
\end{figure}

\begin{figure}[H]
\centering	  
\includegraphics{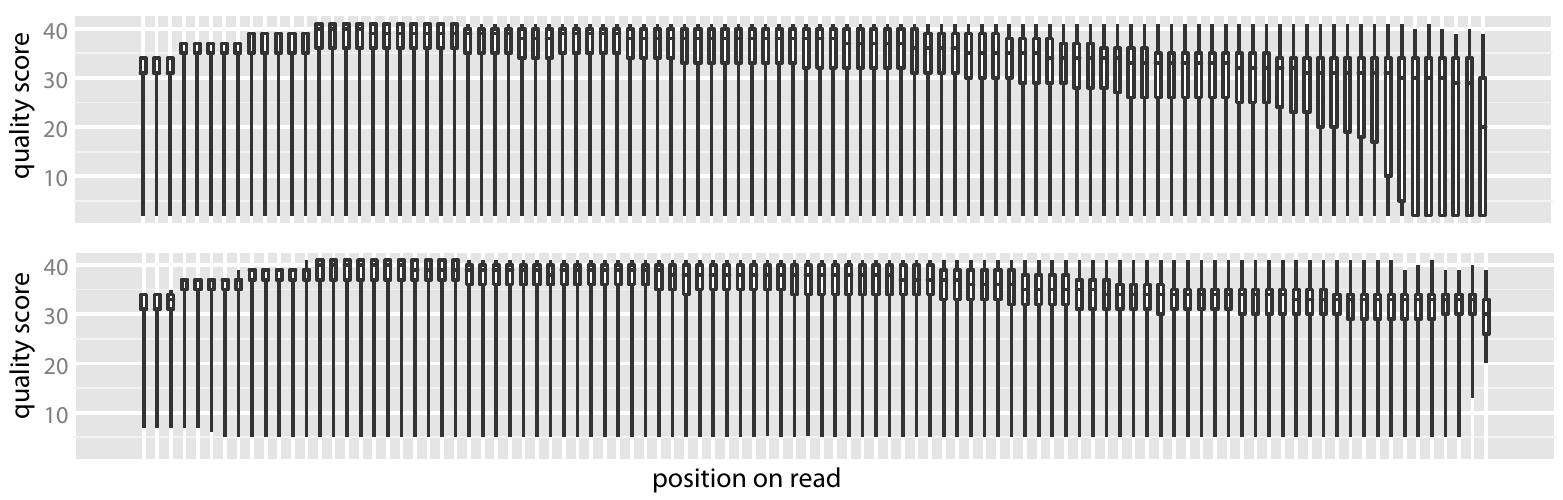}
	\caption{Quality scores in Phred along a read; top graph shows quality prior to cleaning and filtering, bottom shows quality after cleaning.}
\end{figure}

\begin{figure}[H]
\centering	  
\includegraphics{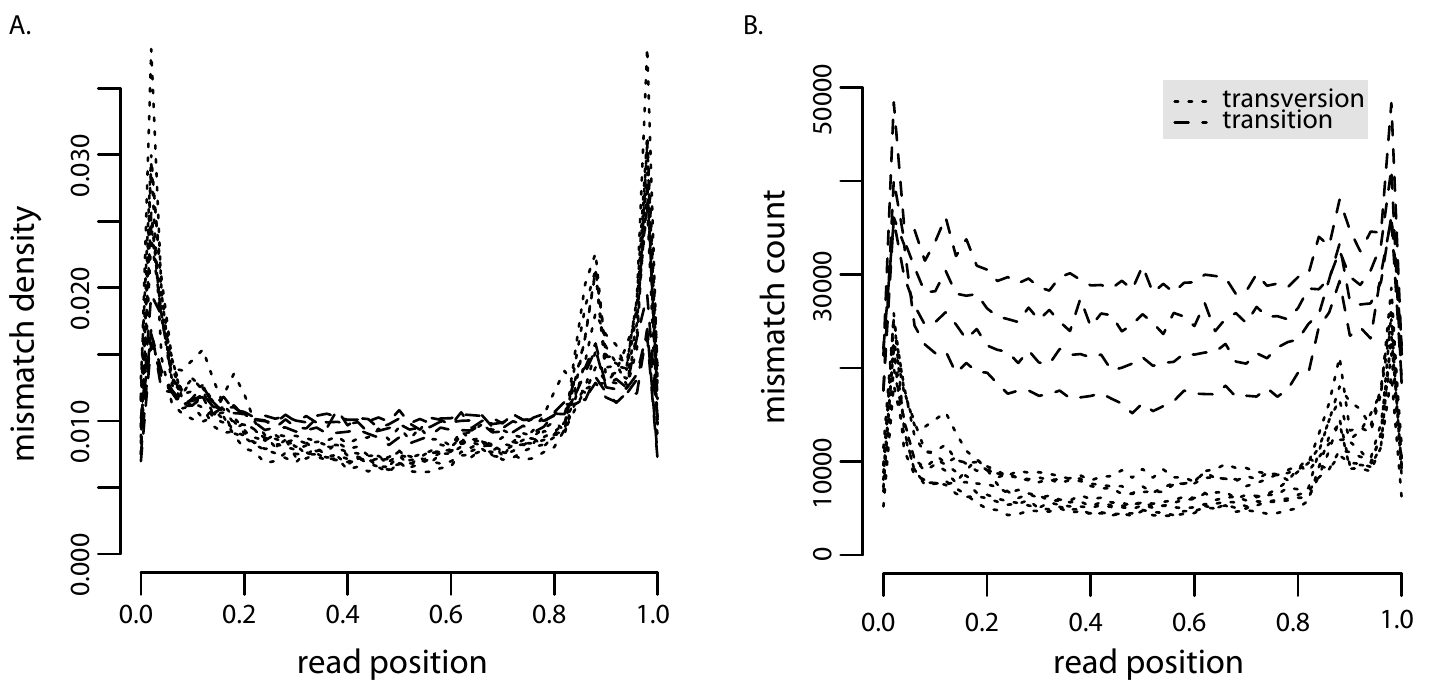}
	\caption{Identified mismatches between reads from a randomly-selected individual and the reference sequence, A. expressed in raw numbers and B. as a density distribution.}
\end{figure}

\begin{figure}[H]
\centering	  
\includegraphics{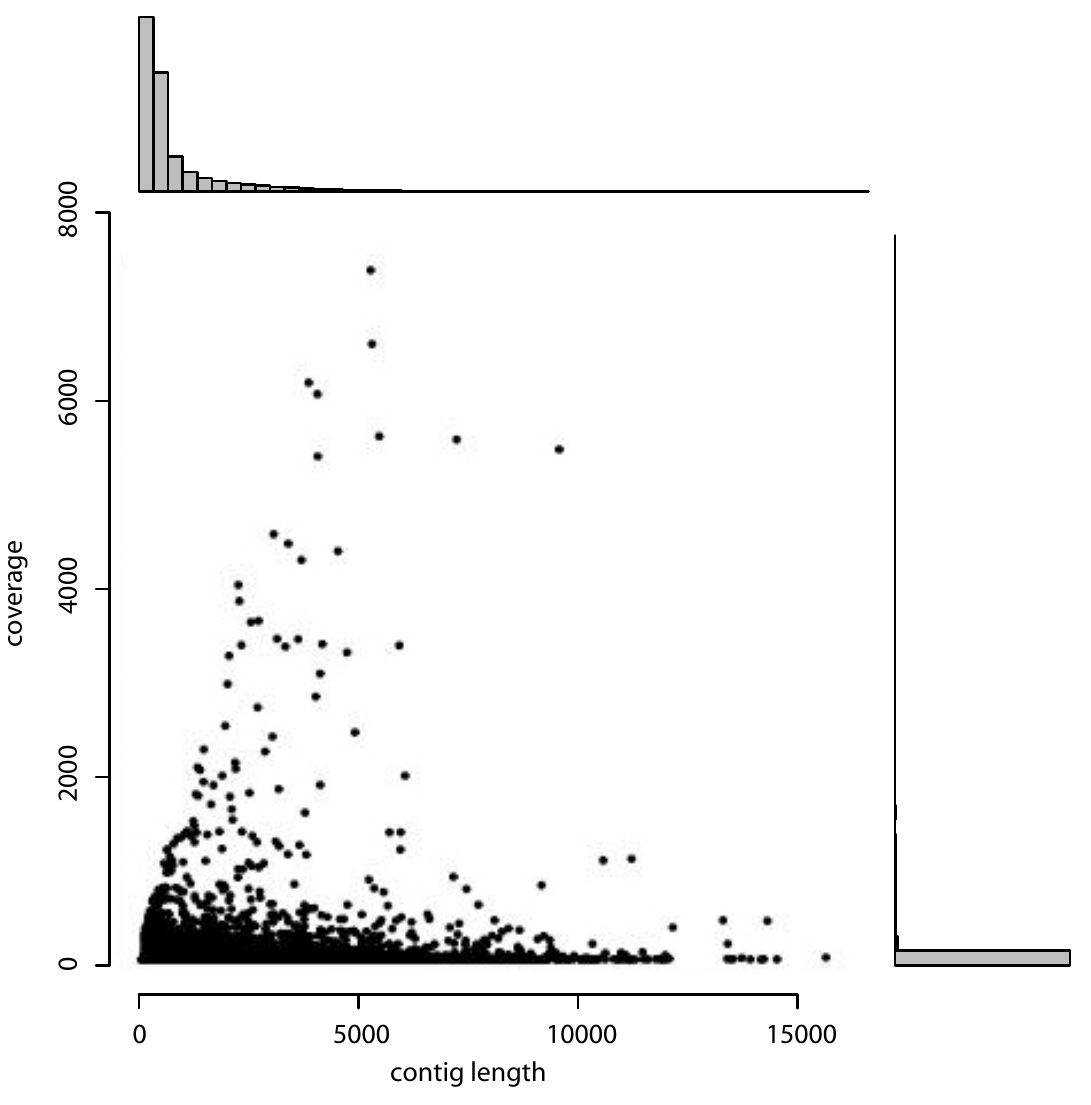}
	\caption{Correlation between contig length and coverage for a randomly-selected final assembly.}
\end{figure}

\begin{figure}[H]
\centering	  
\includegraphics[width=5in]{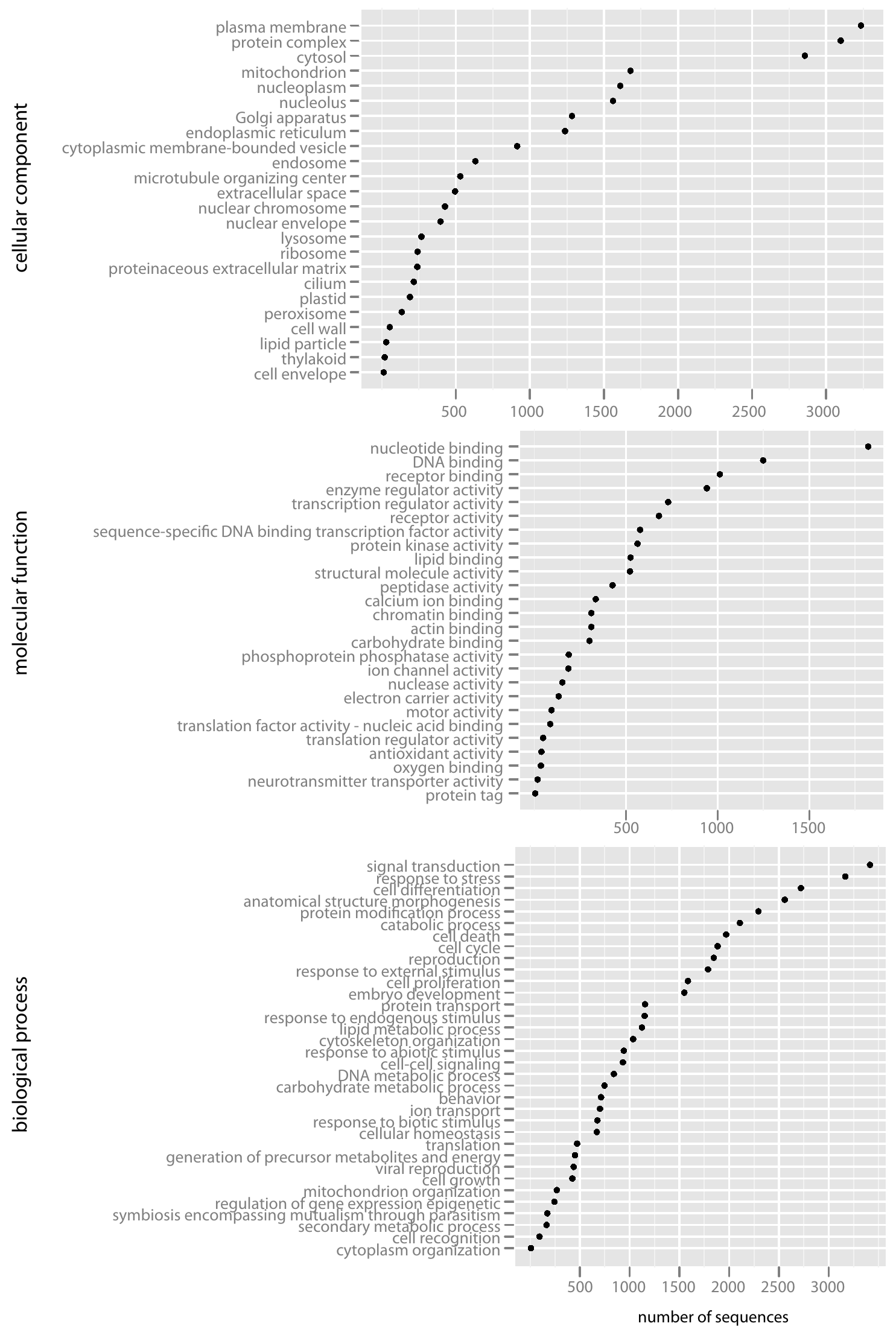}
	\caption{Gene ontology for annotated contigs for a randomly-selected lineage, with respect to cellular component, biological process, and molecular function.}
\end{figure}

\begin{figure}[H]
\centering	  
\includegraphics{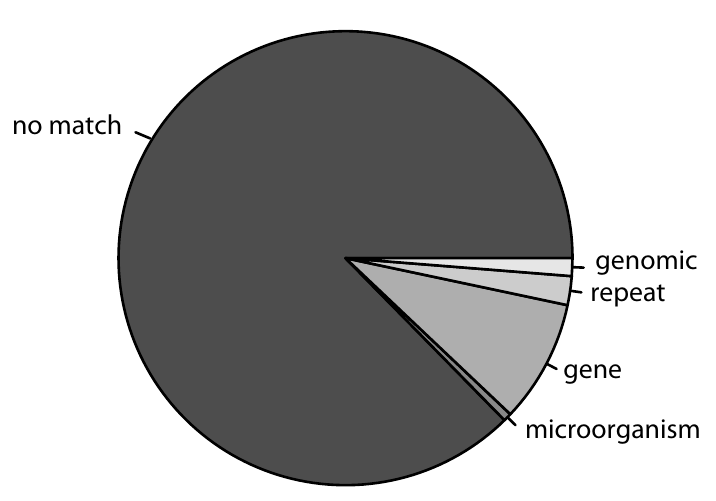}
	\caption{Identify of unannotated contigs from a randomly selected assembly, as identified from a {\sffamily BLAST} search to the {\sffamily NCBI} 'nr' nucleotide database.}
\end{figure}

\begin{figure}[H]
\centering	  
\includegraphics{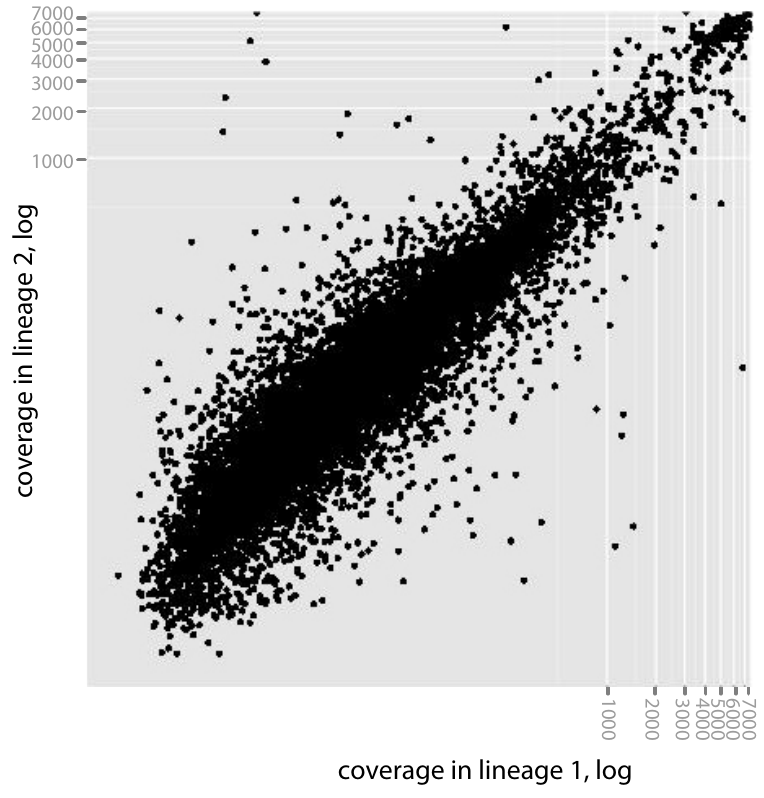}
	\caption{Correlation in coverage between homologous, annotated contigs for a randomly-selected lineage-pair.}
\end{figure}

\begin{figure}[H]
\centering	  
\includegraphics{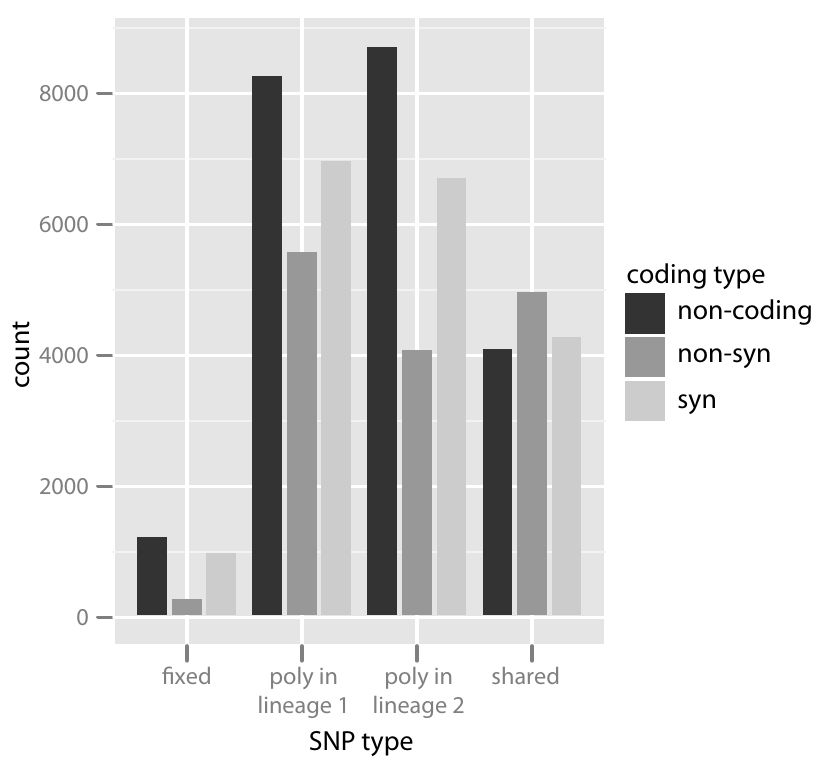}
	\caption{Summary of SNPs found in a randomly-selected lineage-pair, annotated with respect to SNP and coding type.}
\end{figure}

\begin{figure}[H]
\centering	  
\includegraphics{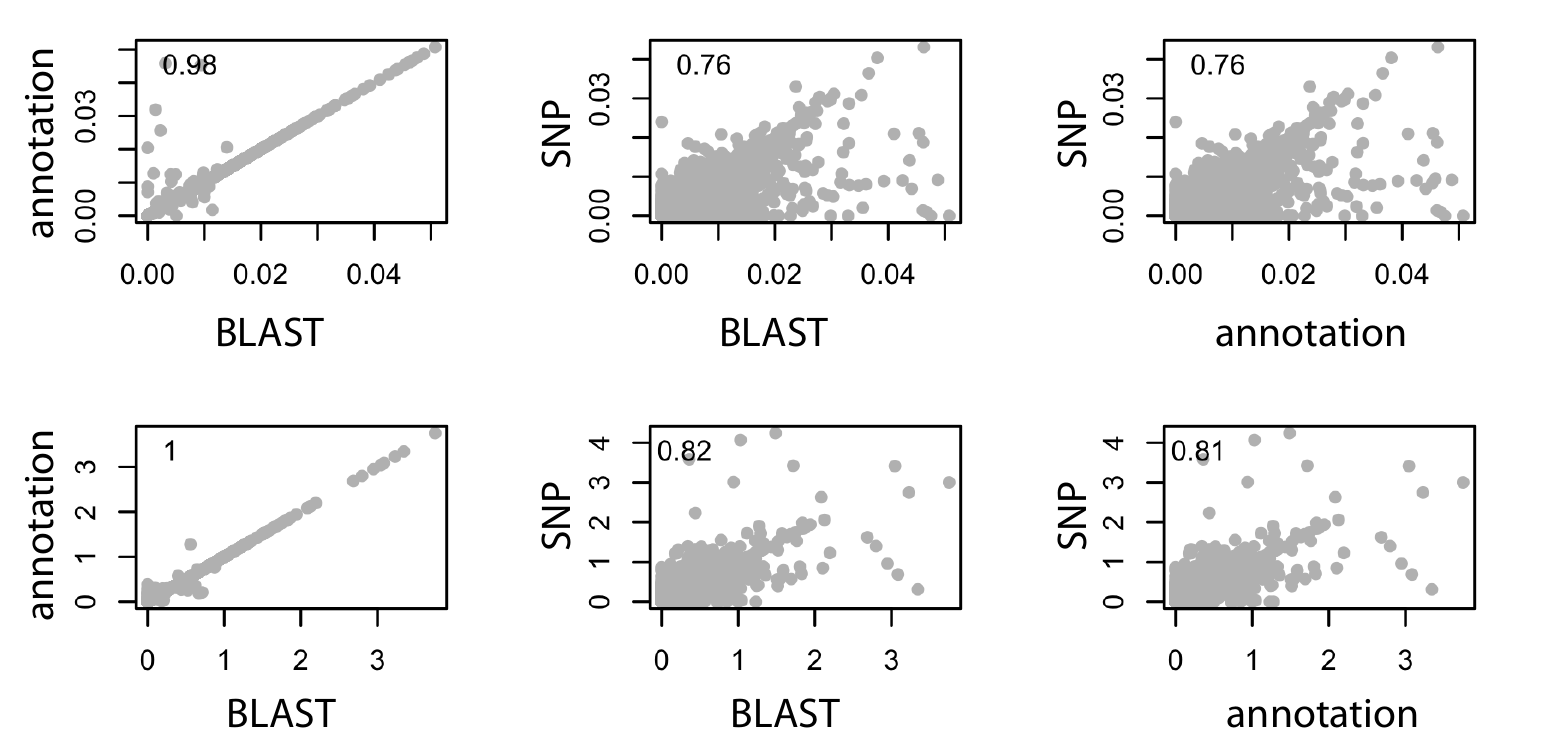}
	\caption{Top row shows correlation in sequence divergence and bottom row shows correlation in inferred $dN/dS$ ratios for homologs for a randomly-selected lineage-pair for three methods of homolog discovery.}
\end{figure}

\end{document}